\documentclass[12pt,a4paper]{article}
\usepackage[T1]{fontenc}
\usepackage[utf8]{inputenc}
\usepackage{booktabs}
\usepackage{amsmath}
\usepackage{graphicx}
\usepackage{hyperref}
\usepackage{xcolor}
\usepackage{siunitx}
\usepackage{lineno}
\usepackage{cite}

\begin{document}
\pagenumbering{gobble}
\title{\bf{Scattered light noise characterisation at the Virgo interferometer with tvf-EMD adaptive algorithm}}

\author{Alessandro Longo$^{1,2}$, Stefano Bianchi$^1$, Wolfango Plastino$^{1,2}$,\\ Nicolas Arnaud$^{3,4}$, Antonino Chiummo$^{3}$, Irene Fiori$^{3}$, \\ Bas Swinkels$^{5}$, Michal Was$^{6}$,
}

\date{\small \it{$^1$Department of Mathematics and Physics, Roma Tre University, Via della Vasca Navale, 00146, Rome, Italy}\\
\it{$^2$INFN, Sezione di Roma Tre, Via della Vasca Navale 84, 00146 Rome, Italy.}\\
\it{$^{3}$ European Gravitational Observatory (EGO), I-56021 Cascina, Pisa, Italy.}\\
\it{$^{4}$ Universit\'{e} Paris-Saclay, CNRS/IN2P3, IJCLab, 91405 Orsay, France.}\\
\it{$^5$ Nikhef, Science Park 105, 1098 XG Amsterdam, The Netherlands.}\\
\it{$^6$ Laboratoire d'Annecy de Physique des Particules (LAPP), Univ. Grenoble Alpes, Universit\'{e} Savoie Mont Blanc, CNRS/IN2P3, F-74941 Annecy, France.}\\ 
}

\maketitle

\begin{abstract}
A methodology of adaptive time series analysis, based on Empirical Mode Decomposition (EMD) and on its time varying version tvf-EMD, has been applied to strain data from the gravitational wave interferometer (IFO) Virgo in order to characterise scattered light noise affecting the sensitivity of the IFO in the detection frequency band. Data taken both during hardware injections, when a part of the IFO is put in oscillation for detector characterisation purposes, and during periods of science mode, when the IFO is fully locked and data are used for the detection of gravitational waves, were analysed. The adaptive nature of the EMD and tvf-EMD algorithms allows them to deal with nonlinear non-stationary data and hence they are particularly suited to characterise scattered light noise, which is intrinsically nonlinear and non-stationary. Results show that tvf-EMD algorithm allows to obtain a more precise outcome compared to the EMD algorithm, yielding higher correlation values with the auxiliary channels identified as the culprits of scattered light noise.
\end{abstract}
\section{Introduction}
\begin{figure}[t!]
\centering
\includegraphics[scale=0.4]{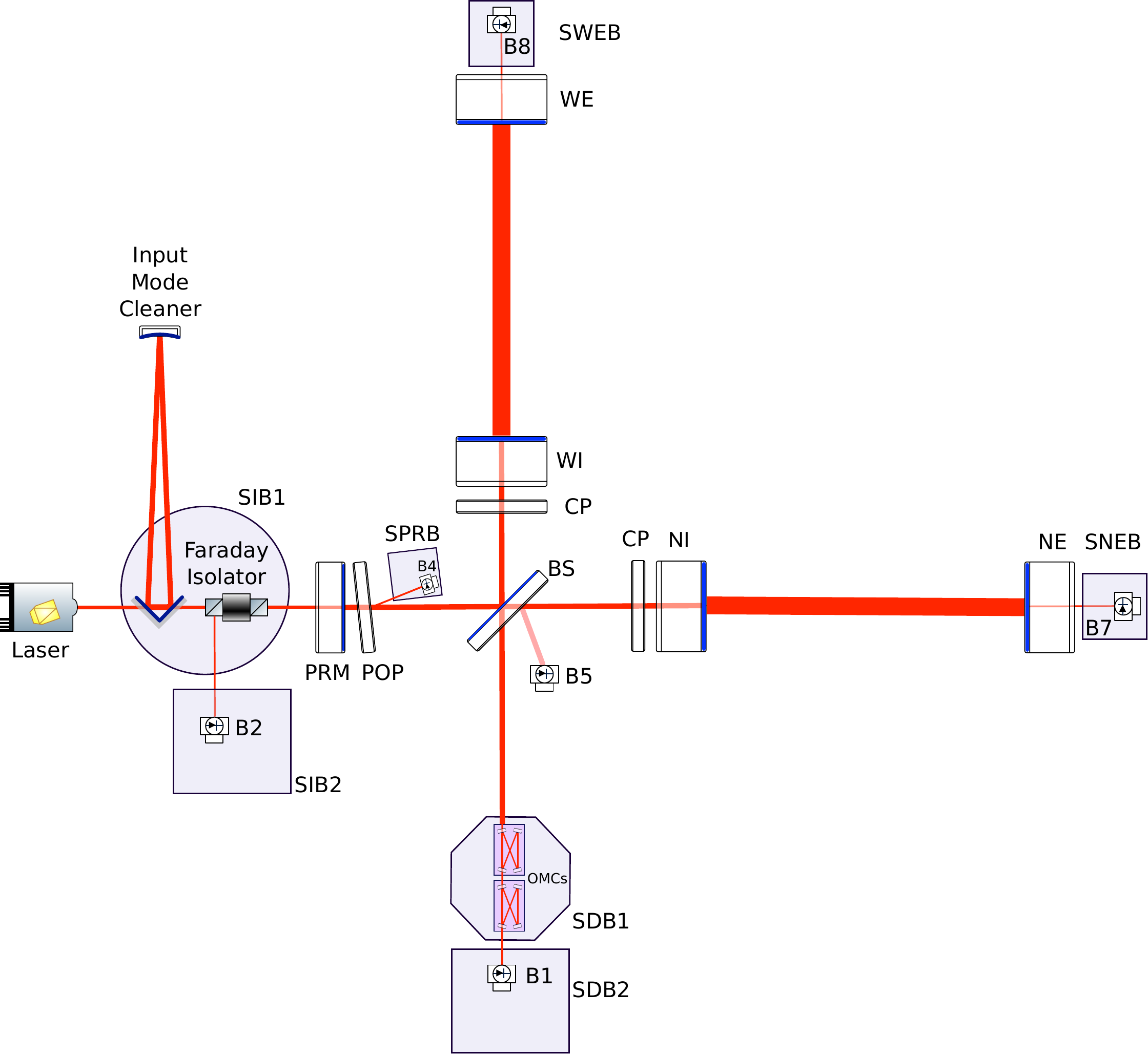}
\caption{Schematics of the Virgo IFO in its advanced configuration. The optical benches are: the laser bench, the suspended injection benches (SIB1-2), the suspended west end (SWEB) and north end bench (SNEB), the suspended power recycling bench (SPRB) and the suspended detection benches (SDB1-2). The mirrors are: the power recycling mirror (PRM), the west input (WI), the north input (NI), west end (WE) and north end (NE) mirrors. Figure also shows the compensation plates (CP) and the pick off plate (POP), the output mode cleaner (OMC) and various photodiodes (B1, B2, B4, B5, B7, B8). Figure adapted from \cite{Acernese_2018}.}
\label{IFO}
\end{figure}

Virgo is an interferometer (IFO) located near Cascina, Pisa (Italy), and together with the laser interferometer gravitational-wave observatory (LIGO), which consists of two IFO located respectively in Hanford and Livingston (United States), it forms a network of three detectors whose goal is to detect gravitational waves generated by events such as the inspiral and merging of two black holes \cite{Acernese_2014,Caron_1997,Aasi_2015}. A schematic of the Virgo IFO in its advanced configuration can be seen in Figure \ref{IFO}, where the suspended injection benches (SIB1-2), the suspended west end bench (SWEB) and the suspended north end bench (SNEB) are visible.
In this paper, a methodology of adaptive time series analysis, based on the one previously implemented in LIGO \cite{Valdes_2017}, is tested on the Virgo strain data in order to characterise sources of scattered light noise. This particular noise can affect the sensitivity of the Virgo IFO in the gravitational wave detection frequency band, ranging from approximately 10 \si{Hz} to a few \si{kHz} \cite{Acernese_2006}. Adaptive methodologies such as Empirical Mode Decomposition (EMD) and Ensemble EMD are widely used in many fields of science, for example in seismology \cite{Wang_2012,Zhang_2003,Chen_2015} and in speech pattern classification \cite{He_2011,Khaldi_2008,Chatlani_2011} while in particular the tvf-EMD algorithm has been already used in seismometer and radionuclide time series analysis \cite{Longo_2020, Longo_2018_emd}.
As described in \cite{Valdes_2017}, besides known features in the sensitivity curve such as peaks due to mechanical resonances or main power harmonics, non-stationary noise can affect the sensitivity of gravitational wave interferometers. Scattered light noise is a relevant example since it affects the sensitivity of the detector in the GW detection frequency range. As described in \cite{Acernese_2020}, to experimentally mitigate scattered light noise during the last observational run, also referred to as O3, a set of baffles has been installed during the upgrade of Virgo to its advanced configuration, in order to intercept and dump most of the stray beams generating scattered light. 
In this paper, the problem of scattered light mitigation is tackled form a data analysis point of view, using adaptive algorithms to quickly identify the scatterer (i.e. reflective surface) that has most likely generated scattered light in the strain data. In a km long IFO, where the possible sources of scattered light are many, this is often a time consuming operation.
Scattered light noise is a phase noise due to interactions between the instrument and the environment and it consists of light exiting the main laser beam, being scattered by moving objects, and then recombining with the main beam \cite{Accadia_2010scatt,Accadia_2010scatt2}. 
After reflecting once from the scattering surface, the scattered light phase angle is given by \cite{Valdes_2017,Accadia_2010scatt}  
\begin{equation}
\phi_{scattering}(t)=2\frac{2\pi}{\lambda}(x_{0}+\delta x_{surface}(t))
\label{scattering}
\end{equation}
where $x_{0}$ is the static optical path measured with respect to some reference system and $\delta x_{surface}(t)$ is
the displacement of the moving surface along the main direction of the beam, also referred to as $z$, while $\lambda=1.064$ \si{\mu m} is the Virgo laser wavelength. 
As described in \cite{Accadia_2010scatt}, the noise introduced by scattered light can be written as
\begin{equation}
h_{sc}(t) \propto \sin( \phi_{scattering}(t))
\end{equation}
Distinctive features of scattered light noise, which is a transient nonlinear noise that changes both in time and frequency domain, are arch-shaped figures appearing in the spectrograms in the detection frequency band, also referred to as \emph{fringes}.
The fringes frequency can be computed from Equation \ref{scattering} taking the time derivative \cite{Valdes_2017,Accadia_2010scatt}
\begin{equation}
f_{fringe}(t)=\frac{1}{2\pi}\phi'_{scattering}(t)=\frac{2}{\lambda}|v_{surface}(t)|
\label{ffringe}
\end{equation}
where $v_{surface}$ is the velocity of the scatterer and the prime symbol stands for time derivative. 
Equation \ref{ffringe} is also referred to as \emph{predictor}, since it gives an insight on the features of scattered light noise appearing in the spectrograms of the differential arm motion (DARM) or of the power recycling cavity length (PRCL) degree of freedom. The PRC is formed between the power recycling mirror (PRM) and the NI and WI mirrors, as can be seen in Figure \ref{IFO}, and is used to recycle the power which gets sent back towards the laser source in order to enhance the circulating power \cite{Acernese_2018}.
If the light is scattered $N$ times before recombining with the main beam, the fringes will appear at higher frequency, given by 
\begin{equation}
f_{fringeN}(t)=Nf_{fringe}(t)
\label{harmonic}
\end{equation}
The methodology adopted in this paper follows the approach described in \cite{Valdes_2017}, where the instantaneous amplitude of modes extracted by the adaptive algorithm EMD is correlated with predictors from many auxiliary channels, using the Pearson correlation coefficient. Results obtained from EMD \cite{Huang_1998,Huang_2014,Wu_2004} and from its recently developed time varying version, tvf-EMD \cite{Li_2017}, are compared in this paper. It is found that the tvf-EMD algorithm gives higher values of correlation in all the tested cases. The paper is organised as follows: in Section \ref{methodology}, the adopted methodology is briefly described; in Section \ref{results}, the results of the analysis are presented and discussed while conclusions are in Section \ref{conclusions}.


\section{Methodology}\label{methodology}
Time series of strain data from the Virgo detector have been analysed in order to identify possible sources of scattered light noise. This has been done making use of adaptive methodologies, such as EMD and tvf-EMD, which are hereafter described.

\subsection{Empirical mode decomposition}\label{EMD}
EMD is an adaptive algorithm first introduced by Huang that allows to deal with nonlinear and non-stationary time series and to extract their oscillatory modes, referred to as intrinsic mode functions (IMFs). To be an IMF, such oscillatory functions must respect the following two conditions:
\begin{itemize}
\item	The number of extrema and zero crossings must be equal or differ at most by one.
\item The mean of the upper and lower envelope must be zero.
\end{itemize}
The procedure to obtain the IMFs is the following: upper and lower extrema contained in the data are fitted with cubic splines and the mean of the upper and lower envelopes is subtracted from the time series data. This procedure is iterated until the two aforementioned conditions are met. This process is referred to as \emph{sifting}. Having obtained the first IMF, it is subtracted from the data and the process is repeated on the remainder of the time series until a slowly varying function $T(t)$ is obtained, which represents either the trend, if present, or the baseline wandering of the data. This way, a given time series $X(t)$ can be represented by the following expansion
\begin{equation}
X(t) = \sum_{j=1}^{K} c_{j}(t) + T(t)
\label{eq:imfs}
\end{equation}
where $c_{j}$ is the $j$th IMF, $t=1 \dots L$, with $L$ being the length of the time series, and $K$ is the number of IMFs that have been extracted by the EMD algorithm. IMFs are mono-component or narrow band oscillatory modes and Hilbert spectral analysis provides physically meaningful estimation of their instantaneous amplitude (IA) and frequency (IF) \cite{Loughlin,Cohen,Boash,Jones_1990}. The IF and IA of a signal $x(t)$ are defined in terms of its analytic signal
\begin{equation}
z(t) = x(t) + iHT[x(t)]=a(t)e^{i\phi(t)}
\label{eq:analytic}
\end{equation}
where $HT[\cdot]$ is the Hilbert transform of $x(t)$ and is defined by 
\begin{equation}
HT[x(t)]=y(t)=\frac{1}{\pi}PV\int_{-\infty}^{\infty}\frac{x(\tau)}{t-\tau}d\tau,
\label{eq:Hilbert}
\end{equation}
where $PV$ stands for principal value integral. The Hilbert transform is the convolution product of $x(t)$ and $1/\pi t$. IA and IF are obtained by means of the following expressions
\begin{equation}
a(t)=\sqrt{x(t)^{2}+y(t)^{2}} \quad ; \quad f_{i}=\frac{1}{2\pi}\phi'(t)
\label{IA_IF}
\end{equation}
The combination of EMD and Hilbert spectral analysis is referred to as Hilbert-Huang spectral analysis and it provides a time frequency ($t$-$f$) representation of the data, known as Hilbert-Huang transform, which has higher resolution compared to the Fourier based methods. 

\subsection{Time varying filter EMD}\label{tvf-EMD}
A known drawback of the EMD application to a noisy dataset is mode mixing, namely an IMF containing oscillations of widely different scales or different IMFs having very similar ones \cite{Huang_1998}. To deal with mode mixing and intermittency related problems and to improve its frequency resolution, a recent modification of the EMD algorithm was introduced, the time-varying filter EMD (tvf-EMD) \cite{Li_2017}. The concept of IMF is replaced by the one of local narrow-band oscillatory modes. To extract local narrow-band signals, B-splines \cite{Unser_1999} are employed as a filter with time-varying frequency cut-off. Furthermore, in the tvf-EMD algorithm the sifting process is stopped when $\theta$, the ratio of the Loughlin instantaneous bandwidth (LIB) and the weighted average IF, is lower than a selected threshold given by the bandwidth threshold ratio parameter $\xi$, as defined in \cite{Li_2017}, Equation 39. For a two-component signal, the LIB is given in terms of instantaneous amplitude and frequency by \cite{Loughlin,Jones_1990}
\begin{equation}
LIB(t)=\sqrt{\frac{a_{1}'^{2}(t)+a_{2}'^{2}(t)}{a_{1}^{2}(t)+a_{2}^{2}(t)}
+\frac{a_{1}^{2}(t)a_{2}^{2}(t)(\phi'_{1}(t)-\phi'_{2}(t))^{2}}{(a_{1}^{2}(t)+a_{2}^{2}(t))^{2}}}
\label{IB_LOUGH}
\end{equation}
The main steps of the sifting procedure, based on the tvf-EMD algorithm, can be found in \cite{Li_2017}, algorithm 3. In the remaining of this paper the term IMFs refers both to the modes extracted by EMD and to the narrow band signals extracted by the tvf-EMD algorithm. 
\subsection{Adopted methodology}
The adopted methodology is based on \cite{Valdes_2017}. Data from the DARM degree of freedom of the IFO, both during period of hardware injections, i.e. when one part of the detector is put in oscillation for detector characterisation purposes, and during science mode, i.e. when the IFO is fully locked and the data acquired are suitable for the search of gravitational waves, were analysed with EMD and tvf-EMD. In one case data from PRCL degree of freedom were also analysed. Results were correlated with predictors computed from 365 auxiliary channels, making use of Equation \ref{ffringe}. The data of the auxiliary channels used for the analysis have been selected from the Virgo interferometer channel database, where the channels having units of \si{\mu m} or \si{\mu rad} and having sampling frequencies of $f_{sampl}=500$ \si{Hz}, $f_{sampl}=200$ \si{Hz} and $f_{sampl}=100$ \si{Hz} have been chosen. The steps of the adopted methodology are the following:
\begin{itemize}
\item Before starting the analysis, data were downsampled to $f_{sampl}=100$ \si{Hz};
\item A lowpass filter with either $f_{cutoff}=30$ \si{Hz}, $f_{cutoff}=15$ \si{Hz} or $f_{cutoff}=10$ \si{Hz} was applied to either DARM or PRCL time series;
\item DARM or PRCL data were decomposed both with EMD and tvf-EMD, obtaining a set of IMFs;
\item These IMFs were normalised to zero mean and unit variance;
\item The IAs of the IMFs were obtained from the magnitude of their analytic signal;
\item Predictors were computed taking the time derivative of the data monitoring $x_{surface}(t)$, i.e. the position of the potential scattering surface with respect to a given reference system. Since a comparison between predictors and the IFs of the obtained modes was not carried out, only the case $N=1$ scattering has been considered.
\item Predictors data were smoothed using a moving average, employing windows of 50 samples;
\item The IA of DARM or PRCL IMFs were correlated, using the Pearson correlation coefficient, with the predictors from various auxiliary channels. The most correlated channel is considered to be the culprit of the scattered light.
\end{itemize}
The relevant parameters of tvf-EMD used during the analysis are the bandwidth threshold ratio $\xi=0.1$, the B-spline order $n=26$, which affects the time varying filter frequency roll-off, and the maximum number of IMFs to be extracted $K=log_{2}L$, rounded upward, where $L$ is the data length.
From a visual inspection of the Hilbert-Huang transform of the data, it has been established that in most cases the first IMF was the one responsible for scattered light noise in the lowpass filtered data. Therefore the first IMF is chosen to compute the correlation with the predictors, unless stated otherwise.


\section{Results and discussion}\label{results}
In this Section, results of the analysis are summarised. Data sampled during hardware injections (HI), during periods in which the detector was in science mode affected by a known source of scattered light, and during the period 09/10/2019 in which the cause of scattered light affecting PRCL degree of freedom was unknown, were analysed. Results are presented in Sections \ref{sec:HI} and \ref{sec:real}, where the most correlated predictor and the IA of the relevant IMF of DARM and PRCL is shown. For a clearer visual comparison, the IAs were smoothed using a moving average with a span of 50 points.

\subsection{Test of the algorithm during hardware injections}\label{sec:HI}
\begin{figure}[!]
\centering
\includegraphics[scale=0.4]{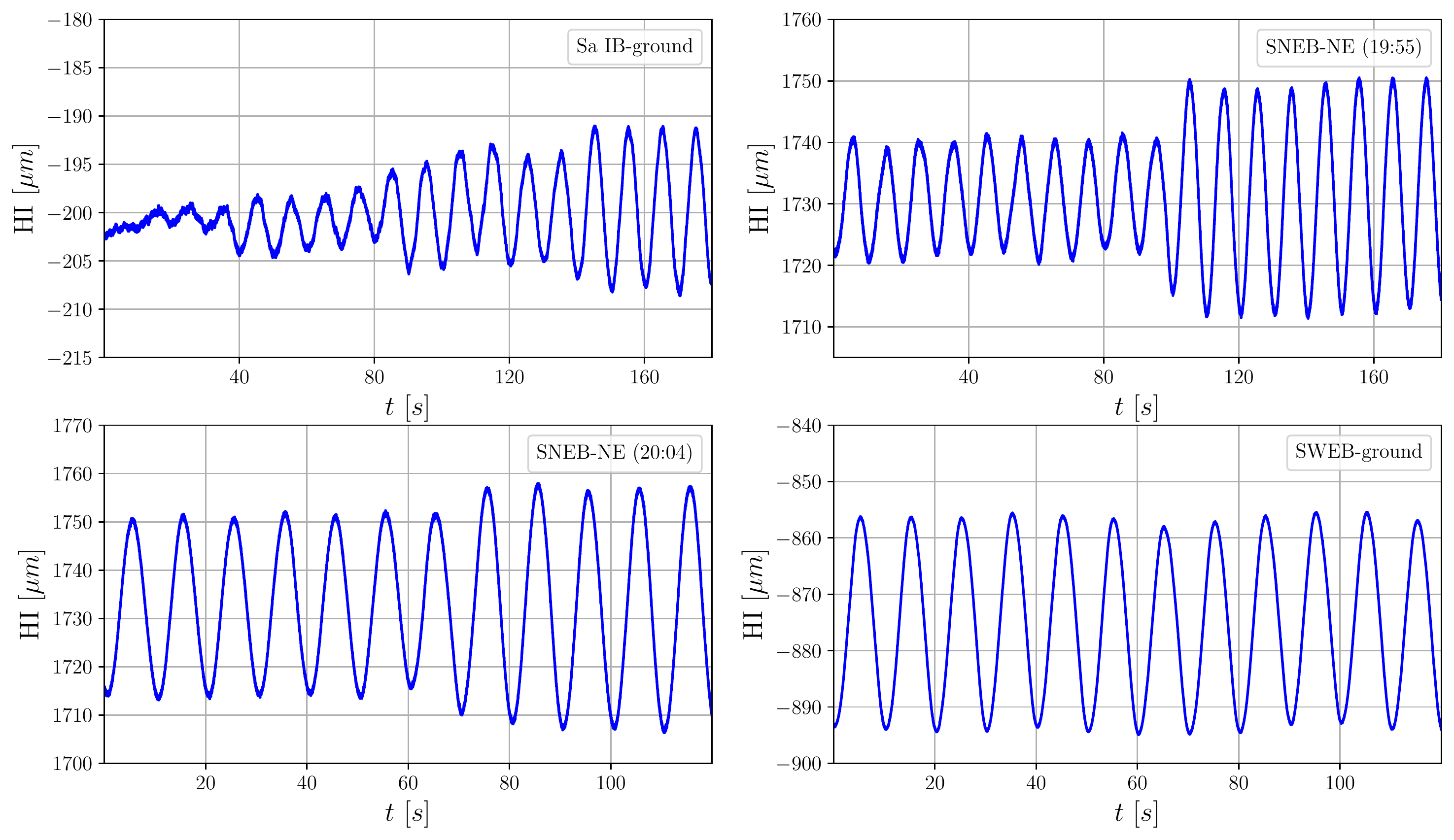}
\caption{Bench movements, w.r.t. the ground, induced by the HI. Lines were injected on the superattenuator of the injection bench (top left panel), on the SNEB (top right and bottom left panels), and on the SWEB (bottom right panel). Sampling frequency is $f_{s}=500$ \si{Hz}.}
\label{fig:HI_time}
\end{figure}

\begin{figure}[t!]
\centering
\includegraphics[scale=0.38]{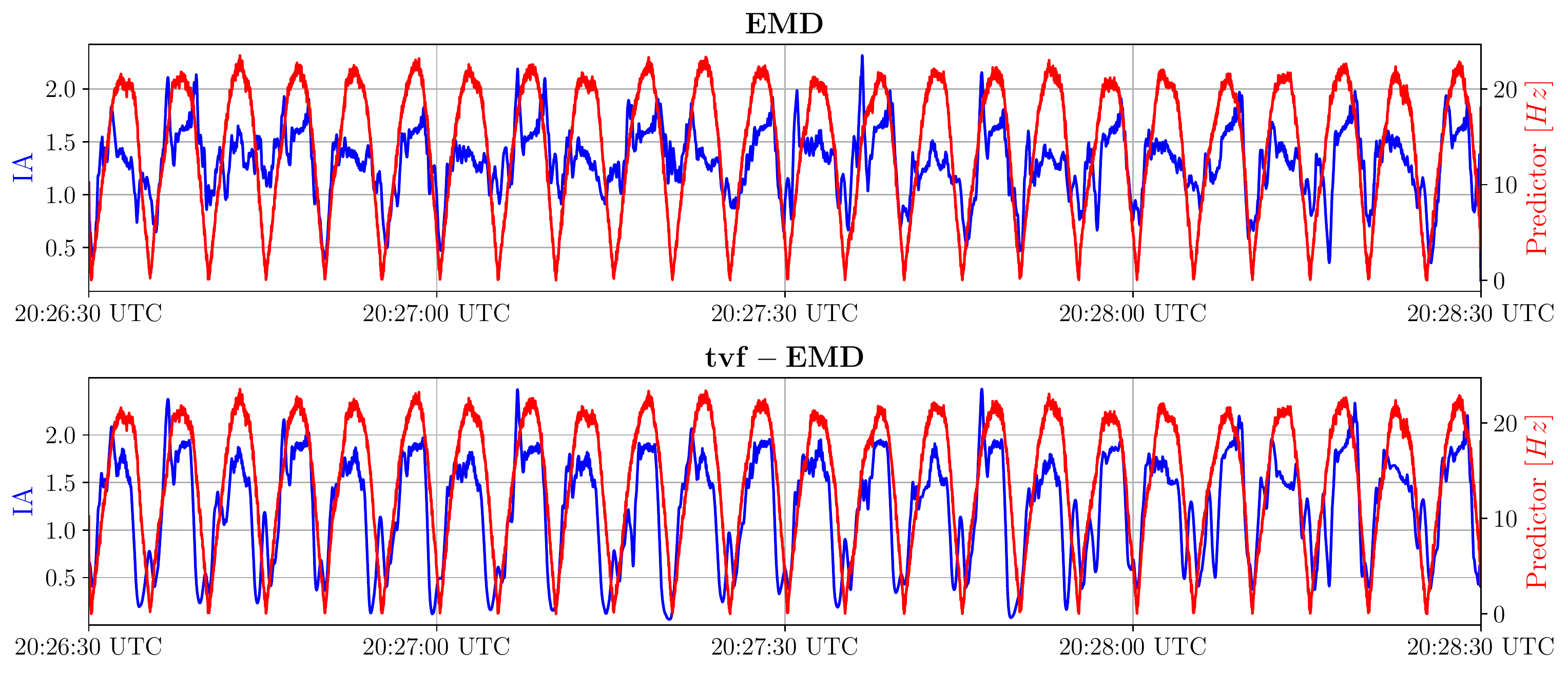}
\caption{Comparison between the predictor (red) for the case of 03/10/2019 - 20:26:30 UTC, and IA (blue) of the first IMF obtained when employing EMD (top panel) and tvf-EMD (bottom panel) to decompose DARM data. The IA obtained by tvf-EMD is physically more meaningful, tracking more precisely the amplitude modulation induced by the HI in the strain data.}
\label{fig:HI_comparison}
\end{figure}

\begin{figure}[t]
\centering
\includegraphics[scale=0.36]{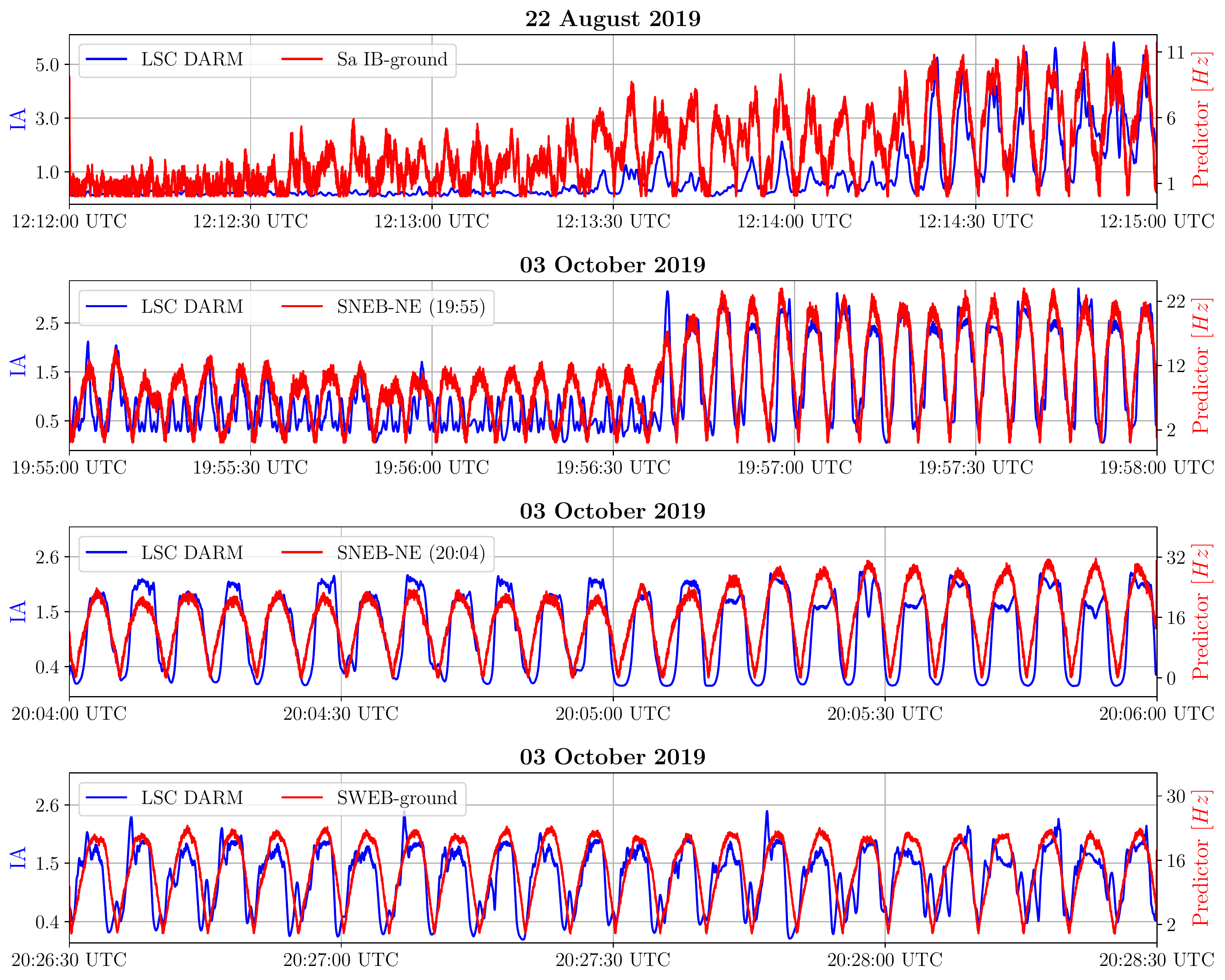}
\caption{Results obtained during controlled HI. The culprit is correctly traced in the four cases to be the superattenuator of the injection bench (Sa IB-ground), the suspended north end bench (SNEB-NE), and the west end bench (SWEB-ground), respectively. Data have been downsampled to $f_{s}=100$ \si{Hz}. The plots are zoomed in the regions of interest for a better visualisation.}
\label{fig:HI}
\end{figure}

The algorithm was initially tested during periods of HI, when lines of known amplitude and frequency are injected in the IFO for detector characterisation purposes, inducing a motion in the optical benches. Hence, during HI it is known what part of the detector is causing scattered light. Four HI were considered and the relevant benches are listed in Table \ref{tab:benches}.
The superattenuator (Sa) is the chain of seismic filters that isolate the Virgo's relevant mirrors such as the test masses, and the term suspended bench refers to the isolation system which isolates the Virgo optical benches from the seismic noise.
The description of such HI can be found in the entries $\#$47091 and $\#46744$ of the Virgo logbook \cite{logbook1} \cite{logbook2}.
In Figure \ref{fig:HI_time} the motions w.r.t. the ground induced by the four different HI, having amplitude up to $20$ \si{\mu m} and frequency of $f_{HI}=0.1$ \si{Hz}, are reported. The top left panel shows the HI on the first seismic filter of the Sa of the injection bench, the top right and the bottom left panels show the HI on the SNEB, while the bottom right panel shows the HI on the SWEB. 
Figure \ref{fig:HI_comparison} shows a comparison between the IA obtained when decomposing DARM data with EMD and tvf-EMD respectively, for the case of 03/10/2019 - 20:26:30 UTC. The IA obtained by tvf-EMD algorithm is more correlated with the predictor obtained for SWEB, as also reported in Table \ref{Tab:unknown}, top panel, tracking more precisely the amplitude modulation induced in the data by the HI.
In Figure \ref{fig:HI}, the results obtained applying the adopted methodology to strain data sampled during the four HI are reported. In blue is the IA of DARM's first IMF (only results obtained with tvf-EMD are shown, due to the fact that tvf-EMD gave higher values of correlation in all cases analysed, see Table \ref{Tab:unknown}) obtained after the normalisation of the IMF to zero mean and unit variance. In red is the predictor for the most correlated channel as obtained by Equation \ref{ffringe}. In all the four cases the algorithm is able to flag the right culprit (the channel of the HI) among all the 365 analysed channels, as can be seen in Figure \ref{fig:HI}. 
Table \ref{Tab:unknown}, top panel, lists the Pearson correlation coefficients, $\rho_{tvf}$ and $\rho_{EMD}$, obtained when decomposing DARM data with tvf-EMD and with EMD, respectively. It can be seen that the tvf-EMD algorithm yields higher values. Table \ref{Tab:unknown} also lists $\rho^{(2)}$, the second best value of correlation obtained for EMD and tvf-EMD. 

\begin{table}[t]\centering
\begin{tabular}{lllp{0.5\columnwidth}}
\toprule
Channel of HI & HI date and time & Culprit ID \\
\midrule
Sa IB-ground & 22 August 2019 - 12:12:00 + 180s (UTC) & Sa IB-ground \\
SNEB LC Z & 03 October 2019 - 19:55:00 + 180s (UTC) & SNEB-NE\\
SNEB LC Z & 03 October 2019 - 20:04:00 + 120s (UTC) & SNEB-NE\\
SWEB LC Z & 03 October 2019 - 20:26:30 + 120s (UTC) & SWEB-ground\\
\bottomrule
\end{tabular}
\caption{Date and times of HI and ID of the obtained culprit. S(W)NEB LC Z stands for bench displacement with regard to (w.r.t.) the ground in the direction of propagation of the beam, Sa IB-ground stands for the position of the SIB1 suspension point w.r.t. the ground (channel ID ``Sa IB F0 X''), SNEB-NE stands for the difference between the position, w.r.t. the ground, of SNEB and the NE building (``SBE SNEB diff bench MIR z'' ), SWEB-ground stands for the difference between SWEB and the ground (``SBE SWEB SA F0 diff LVDT z'') .}
\label{tab:benches}
\end{table}

\begin{table}[b!]\centering
\begin{tabular}{llccccccp{0.5\columnwidth}}
\toprule
Date and time (UTC) & Culprit ID & $\rho_{tvf}$ & $\rho_{EMD}$ & $\rho_{tvf}^{(2)}$ & $\rho_{EMD}^{(2)}$ &$\Delta \rho_{tvf}$ &$\Delta \rho_{EMD}$\\ 
\midrule
\midrule
22/08/2019 - 12:12:00 & Sa IB-ground & 0.73 & 0.73 & 0.24  & 0.26 & 2.04 &1.81\\
03/10/2019 - 19:55:00 & SNEB-NE & 0.80 & 0.69 & 0.28 & 0.14 &1.86 &3.93\\
03/10/2019 - 20:04:00 & SNEB-NE & 0.81 & 0.71 & 0.09 & 0.09 &8.00 &6.89\\
03/10/2019 - 20:26:30 & SWEB-ground & 0.63 & 0.39 & 0.09 & 0.09 &6.00 &3.33\\
\midrule
15/11/2019 - 14:00:00 & SWEB-WE & 0.35 & 0.24 & 0.19 & 0.23 &0.84 &4.30\\
26/11/2019 - 22:00:00 & SWEB-WE & 0.38 & 0.21 & 0.15 & 0.13 &1.53 &0.62\\
27/11/2019 - 04:01:00 & SWEB-WE & 0.48 & 0.32 & 0.10 & 0.17 &3.80 &0.88\\
09/10/2019 - 22:09:40 & EIB-ground & 0.56 & 0.50 & 0.49 & 0.42 & 0.14 & 0.19\\
02/12/2019 - 18:00:00 & SWEB-WE & 0.71 & 0.56 & 0.25 & 0.24 &1.84 &1.33\\
04/12/2019 - 06:00:00 & SWEB-WE & 0.53 & 0.35 & 0.18 & 0.21 &1.94 &0.67\\
12/12/2019 - 02:09:30 & SWEB-WE & 0.74 & 0.49 & 0.32 & 0.17 &1.31 &1.88\\
\bottomrule
\end{tabular}
\caption{Pearson correlation coefficients obtained using tvf-EMD and EMD to decompose DARM or, for the case of 09/10/2019, PRCL time series during HI (top) and during periods affected by scattered light noise with the IFO in science mode (bottom). In all cases tvf-EMD yields higher correlation values. Also reported is $\rho^{(2)}$, the second best value of the correlation obtained among all the other auxiliary channels analysed. These values, for both algorithms, are significantly lower compared to the ones obtained for the culprit. Also shown, for both algorithms, is the relative difference $\Delta \rho =(\rho-\rho^{(2)})/{\rho^{(2)}}$. It can be seen that in most cases tvf-EMD allows a better discrimination of the culprit compared to EMD. 
Regarding the culprit ID, SWEB-WE stands for the difference between the position, w.r.t. the ground, of SWEB and WE building (channel ID ``SBE SWEB diff bench MIR z''), while EIB-ground stands for laser bench (channel ID ``SBE EIB LVDT X'').}
\label{Tab:unknown}
\end{table}

\subsection{Scattered light noise during science mode}\label{sec:real}
\begin{figure}[!t]
\centering
\includegraphics[scale=0.45]{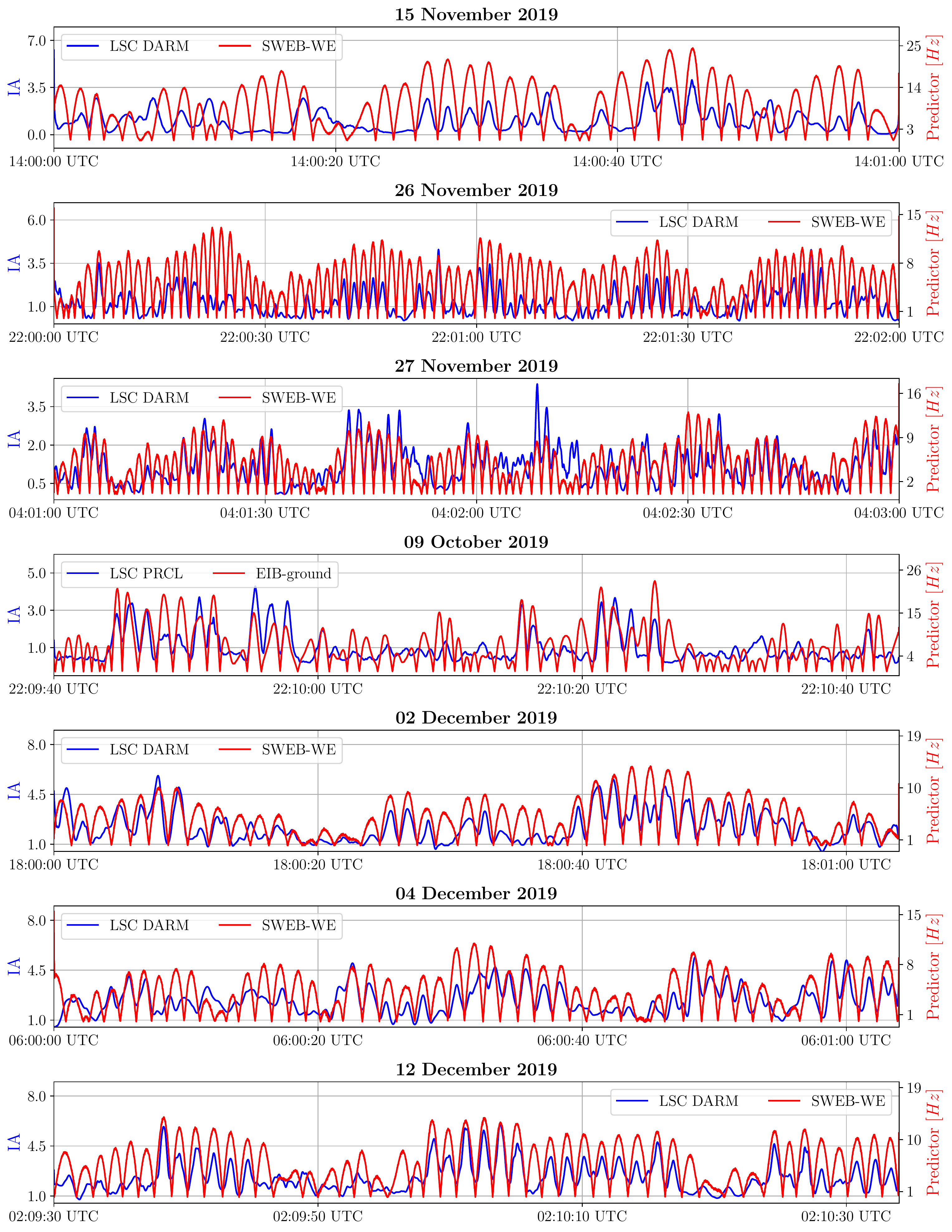}
\caption{Results obtained during periods of high scattered light noise mainly due to the suspended west end bench (SWEB). The culprit is correctly traced back to channels related to the SWEB hosted in the Virgo west end building. In one case the culprit was found to be due to the external injection bench (EIB-ground). Plots are zoomed in the regions of interest for a better visualisation.}
\label{fig:Real}
\end{figure}

\begin{figure}[]
\centering
\includegraphics[scale=0.2]{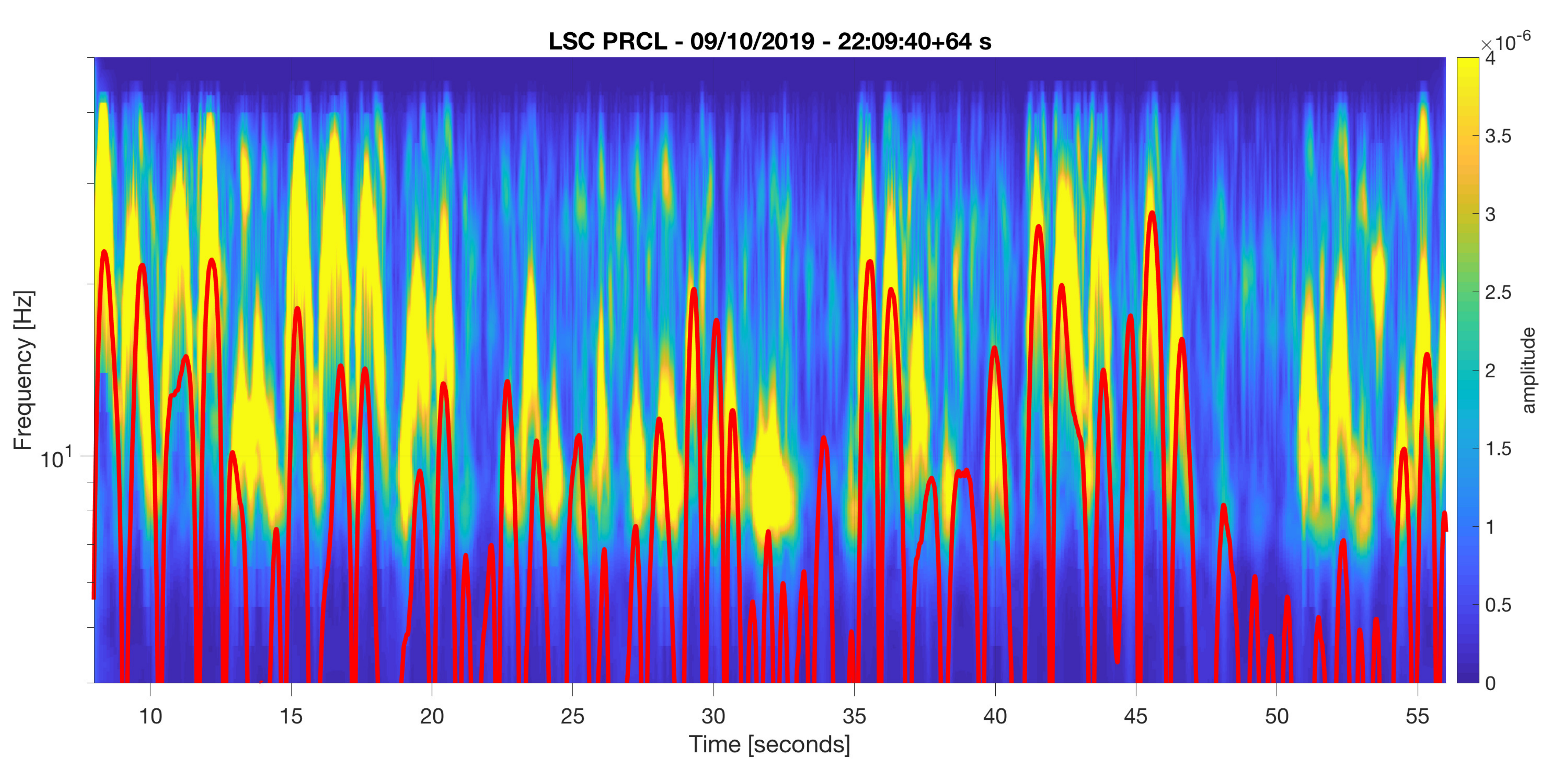}
\qquad\qquad
\includegraphics[scale=0.2]{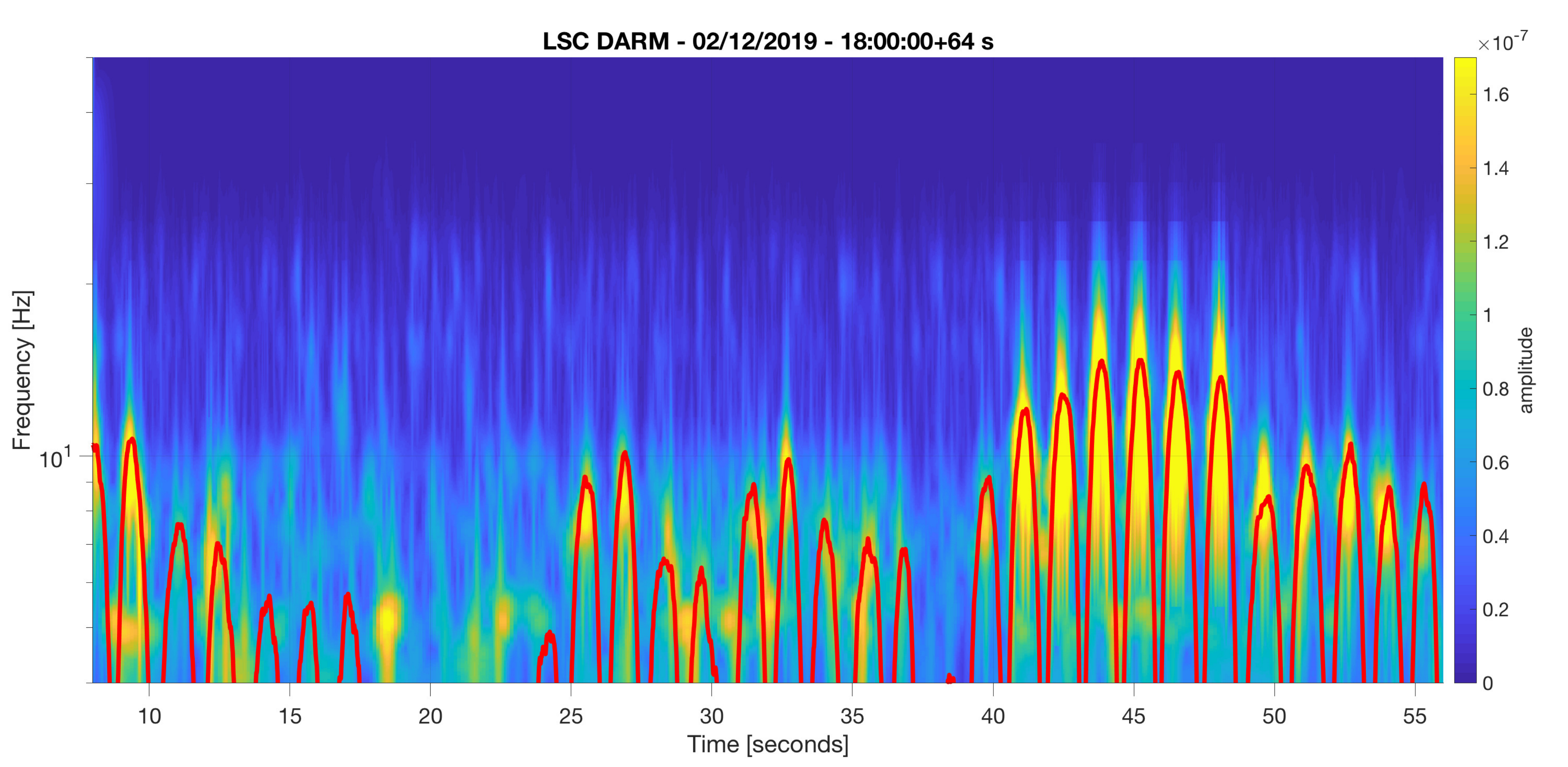}
\qquad\qquad
\includegraphics[scale=0.2]{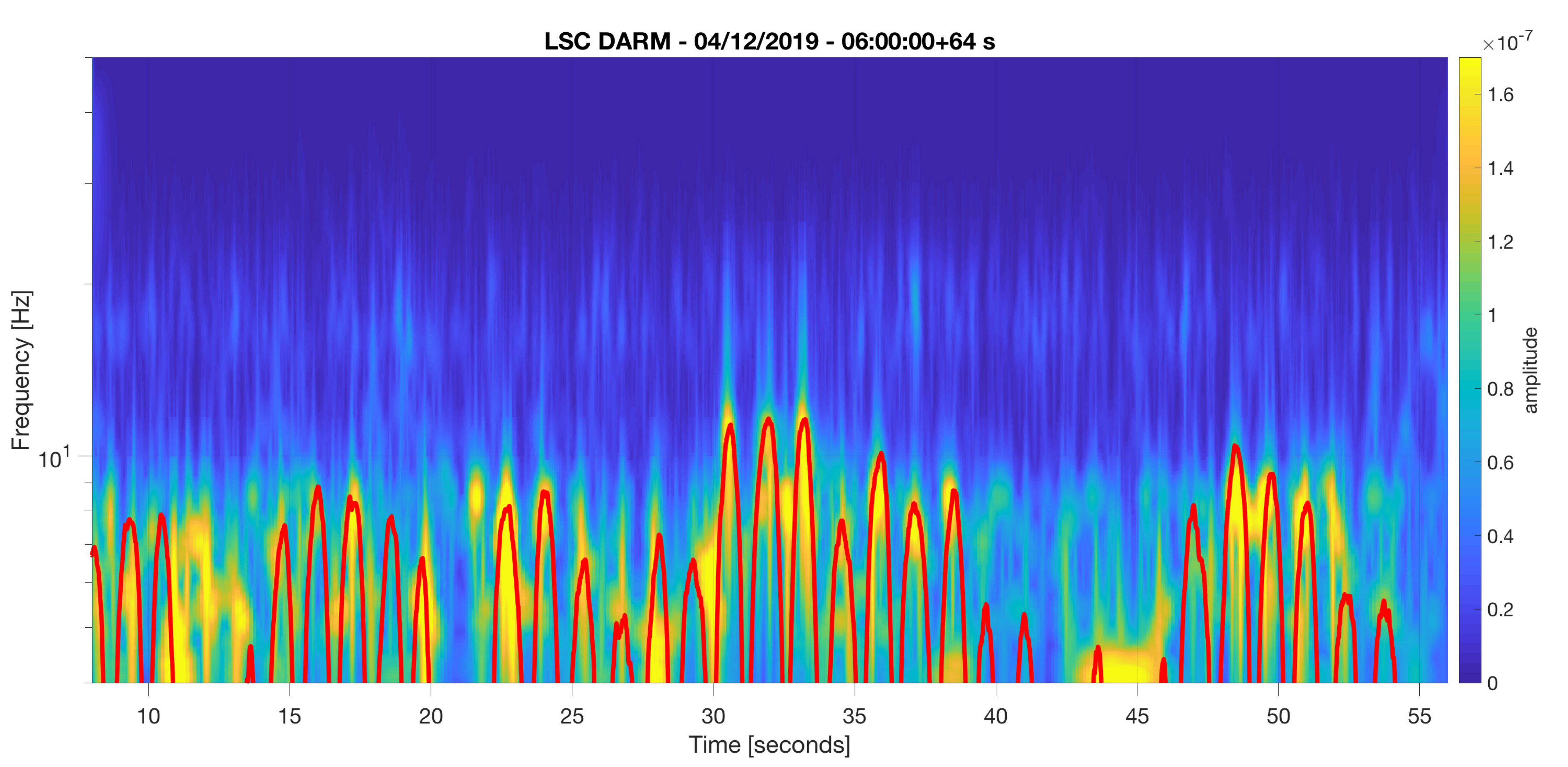}
\qquad\qquad
\includegraphics[scale=0.2]{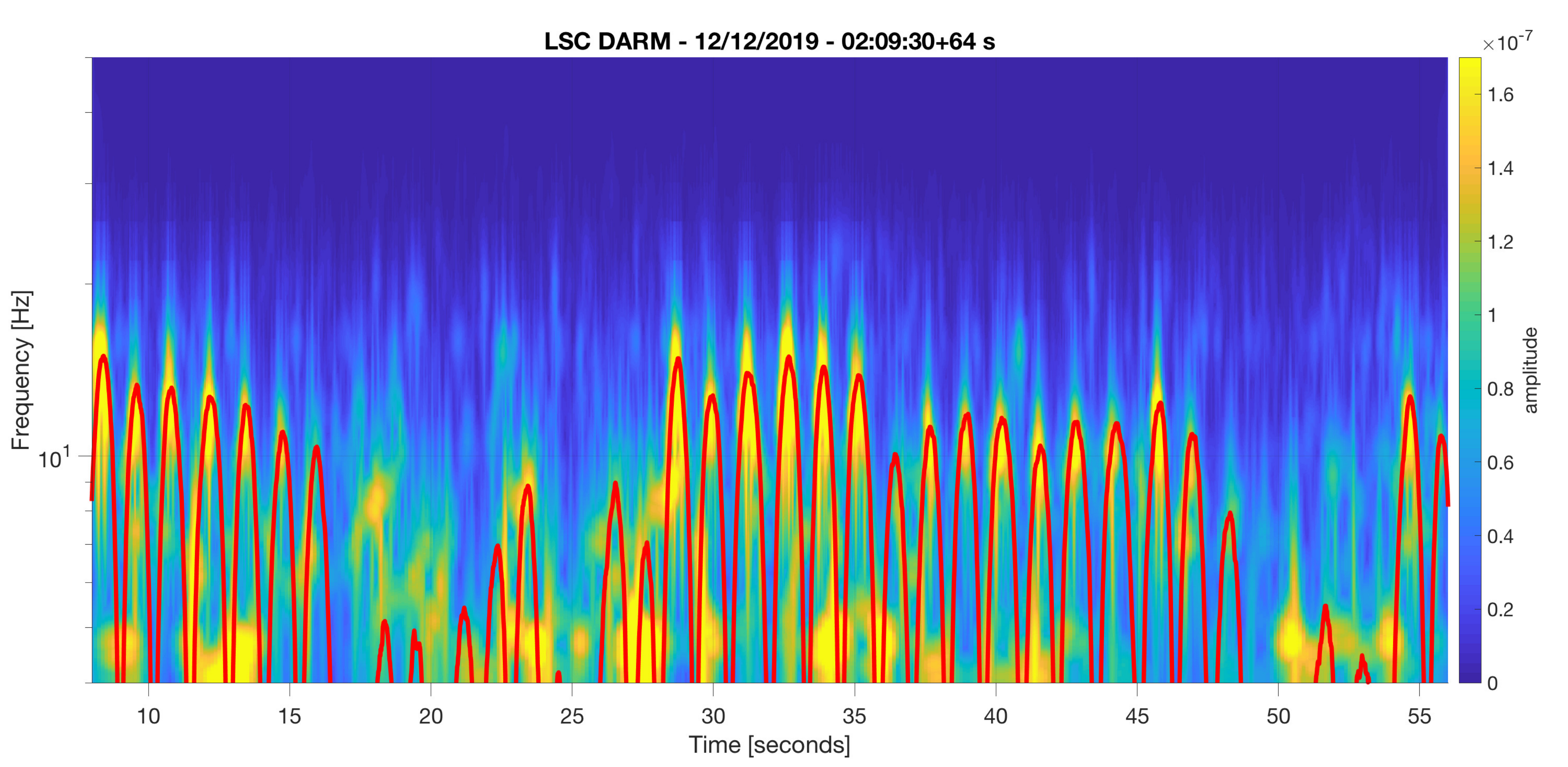}
\caption{Top: Omegagram of PRCL time series. The predictor for the most correlated auxiliary channel, i.e. EIB-ground, is represented in red. Middle and bottom: same as before but for DARM and channel SWEB-WE. The predictors match well the arch shaped fringes appearing in the omegagrams. It should be noted that the scale is different for PRCL and DARM data.}
\label{fig:Omega}
\end{figure}

In this Section, the results obtained from data sampled with the detector in science mode are reported. The data are affected by scattered light noise that is known to be due mainly to the suspended west end bench (SWEB), but also a period in which the origin of scattered light affecting the PRCL degree of freedom is unknown, i.e. the 09/10/2019 one, is considered. The analysed periods are shown in Table \ref{tab:periods}.
Figure \ref{fig:Real} shows the results of the analysis. The IA of the IMFs obtained by decomposing DARM or PRCL data with tvf-EMD and normalised to zero mean and unit variance is shown in blue. The predictor for the most correlated channel, i.e. the culprit of scattered light, is instead shown in red and is obtained by Equation \ref{ffringe}. 
The culprit is correctly found to be a channel related to SWEB, located at the end of the west arm of the IFO, namely SWEB-WE, obtained taking the difference between the position w.r.t. the ground of SWEB and the WE building (channel ID ``SBE SWEB diff bench MIR z'').
For the case of 09/10/2019, scattered light noise was found, in the PRCL degree of freedom, to be due to the external injection bench in EIB-ground, a channel tracking the position of the suspended laser bench w.r.t. the ground (channel ID ``SBE EIB LVDT X'').  
For the case of 02/12/2019 the sum of the IAs of the second and third IMFs has been used to obtain the correlation value of the tvf-EMD case. For the case of 04/12/2019 and 12/12/2019 instead, the reported value of the correlation was obtained summing the IAs of the first and second IMF. This was done since the two considered IMFs have shown, in these cases, the highest values of correlation with the same predictor. Furthermore, such predictor was found to be the culprit also using the Omega algorithm, as can be seen in Figure \ref{fig:Omega}. It should be noted that due to the fact that the cause of scattered light is known, and due to the high coherence with the relevant photodiode, the contribution of scattered light from SWEB could be removed during the procedure of online noise subtraction and did not affect the sensitivity. In Table \ref{Tab:unknown}, bottom panel, the Pearson correlation coefficients obtained for the data taken during science mode are listed. In all cases decomposition performed with tvf-EMD yields better results, i.e. higher correlation values. In Table \ref{Tab:unknown} is also reported the relative difference $\Delta \rho =(\rho-\rho^{(2)})/{\rho^{(2)}}$, for both algorithms, showing that in most cases tvf-EMD allows for a better discrimination of the culprit compared to EMD. Regarding the 03/10/2019 19:55:00 (UTC) and the 12/12/2019 case, for which $\Delta \rho_{EMD}> \Delta \rho_{tvf}$, it should be noted that $\rho^{(2)}$ is in both cases very low, and that $\rho_{tvf}>\rho_{EMD}$. For the 09/10/2019 case the discrimination capability is slightly higher for EMD, but is lower for both algorithms compared to the other cases. 

\begin{table}[!ht]\centering
\begin{tabular}{lp{0.5\columnwidth}}
\toprule
Analysed periods with IFO in science mode \\
\midrule
15/11/2019 - 14:00:00 + 60s (UTC) \\
26/11/2019 - 22:00:00 + 120s (UTC) \\
27/11/2019 - 04:01:00 + 120s (UTC) \\
09/10/2019 - 22:09:40 + 64s (UTC) \\
02/12/2019 - 18:00:00 + 64s (UTC) \\
04/12/2019 - 06:00:00 + 64s (UTC) \\
12/12/2019 - 02:09:30 + 64s (UTC) \\
\bottomrule
\end{tabular}
\caption{Date and time of the analysed data with the IFO in science mode.}
\label{tab:periods}
\end{table}

\clearpage
\hspace{-0.6cm}To validate the obtained results the Omega algorithm was also employed \cite{Graef_2011,Chatterji_2005}. Figure \ref{fig:Omega} shows a comparison between a spectrogram obtained with the Omega algorithm, i.e. Omegagram, and the predictor for the most correlated channel, which is EIB-ground for the top panel and SWEB-WE for the other cases.

\section{Conclusions}\label{conclusions}
In this paper, data from the Virgo IFO have been analysed and characterised using adaptive algorithms. Different periods affected by scattered light noise were considered: periods of controlled HI, periods of science mode where the origin of the noise was known and its effect was subtracted from the gravitational wave strain signal, and a period of scattered light of unknown origin affecting the PRCL degree of freedom of the IFO. Strain data were decomposed both with EMD and tvf-EMD and the IAs obtained from the IMFs were correlated with predictors computed for the 365 auxiliary channels the methodology was tested on. After low-passing, the first IMF was found to be the one relevant for scattered light noise in most cases. The most correlated channel is considered to be the culprit of scattered light. The adaptive algorithms EMD and tvf-EMD were suitable for this analysis since they allow characterisation of data which are both nonlinear and non-stationary, as is the case for scattered light noise. Adaptive algorithms do not make any \emph{a priori} assumption about the expansion basis, which is instead obtained from the data and is given in terms of amplitude and frequency modulated IMFs. The recently developed tvf-EMD is an extension of EMD, improving its frequency resolution while mitigating end effects, mode mixing, and intermittency. The tvf-EMD algorithm yielded higher values of correlation for all the analysed cases compared to standard EMD. 
   
\section*{Acknowledgments}
Data shown in the paper were obtained using the Advanced Virgo monitoring system. We acknowledge the Italian Istituto Nazionale di Fisica Nucleare (INFN), the French Centre National de la Recherche Scientifique (CNRS) and the Foundation for Fundamental Research on Matter supported by the Netherlands Organisation for Scientific Research, for the construction and operation of the
Virgo detector and the creation and support of the EGO consortium.

\clearpage

\section*{Acronyms}
\quad
\begin{table}[!h]\centering
\begin{tabular}{llp{0.5\columnwidth}}
EMD & Empirical Mode Decomposition \\
IFO & Interferometer \\
LIGO & Laser Interferometer Gravitational-wave Observatory \\
SIB1-2 & Suspended Injection Benches \\
SWEB & Suspended West End Bench \\
SNEB & Suspended North End Bench \\
EEMD & Ensemble EMD \\
DARM & Differential Arm Motion \\
PRCL & Power Recycling Cavity Length \\
PRM & Power Recycling Mirror \\
BS & Beam Splitter \\
SPRB & Suspended Power Recycling Bench \\
SDB1-2 & Suspended Detection Benches \\
WI & West Input \\
NI & North Input \\
WE & West End \\
NE & North End \\
CP & Compensation Plates \\
POP & Pick Off Plate \\
OMC & Output Mode Cleaner \\
IMFs & Intrinsic Mode Functions \\
IA & Instantaneous Amplitude \\
IF & Instantaneous Frequency \\
LIB & Loughlin Instantaneous Bandwidth \\
HI & Hardware Injection \\
Sa & Superattenuator \\
LVDT & Linear Variable Differential Transformer \\
EIB & External Injection Bench \\
\end{tabular}
\end{table}
\bibliography{bibl}
\bibliographystyle{ieeetr}

\end{document}